\documentclass[12pt]{article}

\usepackage{latexsym}
\usepackage{graphicx}
\title{\textbf{Cosmic acceleration in a model of scalar-tensor gravitation}}
\author{Sanil Unnikrishnan$^\dagger$  and T. R. Seshadri $\ddagger$ \\
{}\\
{\it Department of Physics \& Astrophysics,}\\
{\it University of Delhi, Delhi 110007,India.}\\
$\dagger${\small e-mail : sanil@physics.du.ac.in}   \\
$\ddagger${\small e-mail :~~ trs@physics.du.ac.in}}

\usepackage{epsfig}
\begin{document}

\maketitle
\begin{abstract}
In this paper we consider a  model of scalar-tensor theory of gravitation in
which the scalar field, $\phi$ determines the gravitational coupling G and
has a Lagrangian of the form,
$\mathcal{L}_{\phi} =-V(\phi)\sqrt{1 - \partial_{\mu}\phi\partial^{\mu}\phi}$.
We study the cosmological consequence of this theory in the matter dominated
era and show that this leads to a transition from an initial decelerated
expansion to an accelerated expansion phase at the present epoch. Using
observational constraints, we see that the effective equation of state today
for the scalar field turns out to be $p_{\phi}=w_{\phi}{\rho}_{\phi}$,
with $w_{\phi}=-0.88$ and that the transition to an accelerated phase happened
at a redshift of about 0.3.
\end{abstract}

\section{Introduction}
Various observations  strongly indicate that at the present epoch
the universe is in a phase of accelerated expansion \cite{riess,perl}.
This accelerated expansion of the universe might be caused by
a non-zero value of the cosmological constant
\cite{wein,paddy,peebles}, or  an evolving scalar field such as
quintessence \cite{ferreria,batra}, phantom field
 \cite{carroll,kaplinghat,cladwell,sami},
tachyon field \cite{paddy1,bagla,copeland} etc.
For a detailed review of dark energy see for example Ref.\cite{DEreview}.

Quintessence refers to a scalar field whose Lagrangian is of the form
\begin{equation}
\mathcal{L}_{\phi} =
\frac{1}{2}\partial_{\mu}\phi\partial^{\mu}\phi -V(\phi)
\end{equation}
In this paper we investigate the consequences of a scalar field whose
Lagrangian is given by,
\begin{equation}
\mathcal{L}_{\phi} =
-V(\phi)\sqrt{1 - \partial_{\mu}\phi\partial^{\mu}\phi}
\end{equation}
Without going into the details
of the fundamental origin of this form, we use this form of the action
at a phenomenological level. It is, however,
appropriate to mention that
this form of the Lagrangian arises from string theory. A detailed investigation of this form of Lagrangian in the context of Einstein-Hilbert action has been studied in Ref\cite{copeland}. Quite
independently (as described in \cite{paddy2}) the structure of this
Lagrangian can also be motivated from the fact that its form is very similar to
that of a relativistic particle in classical mechanics
($\mathcal{L} = -m\sqrt{1 - \dot{q}^{2}}$).

Most of these scalar field models of Dark energy are discussed in the context
of Einstein-Hilbert action in general relativity.
A well known alternative to Einstein's theory is
the Brans Dicke theory \cite{brans} where Newton's gravitational constant $G$ is
identified with the inverse
of a scalar field whose dynamics is governed by the matter
distribution.
The Brans Dicke theory is motivated from Mach's
Principle which in a way suggest that $G$ should be function of epoch
\cite{brans,narlikar,inverno}. Brans Dicke theory can be considered as a special
case of more general scalar tensor theory of gravitation. Accelerated expansion
of the universe in scalar tensor theories of gravitation has been studied by various
authors \cite{peri,elizalde,diaz,sen,farese,boisseau,bartolo,capozziello}.
These authors, however, considered a scalar field Lagrangian which
has a standard form
$ \mathcal{L}_{\phi} =  \frac{1}{2}\partial_{\mu}\phi\partial^{\mu}\phi -V(\phi)$.
In this paper we consider a scalar field which plays a role similar to the
Brans-Dicke field but with a Lagrangian of the form
$\mathcal{L}_{\phi} = -V(\phi)\sqrt{1 - \partial_{\mu}\phi\partial^{\mu}\phi}$.

Throughout
this paper we use the units $\hbar = c = 1$.
We consider the case when $G = \phi ^{2}$ and
$ V(\phi) = \lambda \phi ^{-4}$ where $\lambda$
is a dimensionless parameter (in the units $\hbar = c = 1$).
As pointed out earlier, although this form of the action arises in certain
models of string theory\cite{Piao,PC}, we use this action at a purely phenomenological level,
without analyzing the details of the fundamental origin of this form of the
action.

The precise form of the action and the
resulting  field equations are given in section \ref{action}.
We discuss in section \ref{cc} the
cosmological consequences of this theory in the matter dominated
era and show that this leads to an accelerated expansion of the universe at the
present epoch. The conclusions of this analysis are summarized in section
\ref{conc}.
\section{Action and the field equations } \label{action}
In this paper we consider the action for the scalar tensor gravitation of the
form:
\begin{equation}
\mathcal{A} = \frac{1}{16\pi} \int \left[-\frac{R}{\phi^{2}} -
\frac{\lambda}{\phi^{4}}\sqrt{1 - \partial_{\mu}\phi\partial^{\mu}\phi} +
  16\pi\mathcal{L}_{m}\right] \sqrt{-g}~d^{4}x  \label{STTGtc0}
\end{equation}
We may note the following features of this action. As pointed out earlier,
the scalar field $\phi$ plays a role similar to the Brans-Dicke field
$\phi_{BD}$ in the
standard Brans-Dicke theory. In the latter, the value of $G$ is determined by
$\phi_{BD}^{-1}$. In our model the value of $G$ is determined by $\phi^2$.
Further, the 'kinetic energy' part of $\phi$ in the second term in the action,
is of a form different from the standard Brans-Dicke case.
The function $V(\phi)$ has the dimensions of energy density and
in the units we use here, ($\hbar = c = 1$), it has the units of $M^4$.
Further, in these units $\phi$ has dimensions of $M^{-1}$. With these facts in
mind, we consider a simple form of $V(\phi)$ given by,
\begin{equation}
V(\phi) = \lambda \phi^{-4}
\end{equation}
where, $\lambda$ is a dimensionless parameter.
Varying the action given in Eq. (\ref{STTGtc0}), the Einstein's equation and
the equation of motion for the field are:
\begin{eqnarray}
G_{ab} &=& 8\pi\phi^{2}T_{ab} - \frac{2}{\phi}\left[\phi_{a;b} - g_{ab} \Box{\phi} \right] + \frac{6}{\phi^2}\left[\phi_{a}\phi_{b}  -
  g_{ab}\phi_{\mu}\phi^{\mu}\right]\nonumber \\
& &+ \frac{\lambda}{2\phi^{2}}\left[\frac{\phi_{a}\phi_{b}}{\sqrt{1 -
  \phi_{\mu}\phi^{\mu}}}  +  g_{ab}\sqrt{1 -
  \phi_{\mu}\phi^{\mu}}\right]\label{feqgab}
\end{eqnarray}
\begin{equation}
R = \frac{2\lambda}{\phi^{2}}\frac{1}{\sqrt{1 -
  \phi_{\mu}\phi^{\mu}}} + \frac{\lambda
  \Box{\phi}}{2\phi\sqrt{1 - \phi_{\mu}\phi^{\mu}}} +
  \frac{\lambda \phi^{\mu}(\phi_{i}\phi^{i})_{\mu}}{4\phi(1 -
  \phi_{\mu}\phi^{\mu})^{3/2}}  \label{feqsi}
\end{equation}
where $T_{ab}$ is the energy momentum tensor of the matter. $ T_{ab}$ is defined
as :
\begin{equation}
T^{ab} = -\frac{1}{2\sqrt{-g}}\frac{\delta}{\delta
  g_{ab}}(\mathcal{L}_{m}\sqrt{-g})
\end{equation}
The two field equations Eqs. (\ref{feqgab}) and (\ref{feqsi}) can be combined as:
\begin{eqnarray}
8\pi\phi^{2}T &=&  -\frac{6\Box{\phi}}{\phi} +
\frac{18\phi_{\mu}\phi^{\mu}}{\phi^{2}}
+ \frac{3\lambda}{2\phi^{2}}\frac{\phi_{\mu}\phi^{\mu}}{\sqrt{1 -
    \phi_{\mu}\phi^{\mu}}}\nonumber \\
 & & - \frac{\lambda}{2\phi}\frac{\Box{\phi}}{\sqrt{1 -
    \phi_{\mu}\phi^{\mu}}} -
  \frac{\lambda}{4\phi}\frac{\phi^{\mu}(\phi_{a}\phi^{a})_{,\mu}}{(1 -
    \phi_{\mu}\phi^{\mu})^{3/2}}\label{tsi}
\end{eqnarray}
This equation [Eq. (\ref{tsi})] shows that the evolution of $\phi$ is determined by
matter distribution (through its trace $T$) just as in Brans Dicke Theory but
here the equation is not in as simple form as in Brans Dicke Theory.

In this theory we have one dimensionless parameter $\lambda$ whose value must
be fixed observationally. We will
fix its value by considering the cosmological evolution determined by this
theory and comparing it with the observed parameters for the universe.
\section{Cosmological Consequences}  \label{cc}
The field $\phi$, in general, will be a function of space and
time \textit{i.e} $\phi = \phi(\vec{x},t)$. We, however, will consider
the field to be homogeneous and isotropic on cosmic scales.
Hence, the field $\phi = \phi(t)$.
We will consider spatially flat Friedmann Robertson Walker metric of the form:
\begin{equation}
ds^{2} = dt^{2} - a(t)^{2}[dr^{2} + r^{2}d\theta ^{2} +
  r^{2}\sin{\theta}^{2}d\phi^{2}]\label{lineE}
\end{equation}
Here $a(t)$ is the scale factor. The Friedmann equations governing the
evolution of $a(t)$ are given by
\begin{eqnarray}
\left(\frac{\dot{a}}{a}\right)^{2} &=& \frac{8\pi\phi^{2}}{3}\rho +
 2\frac{\dot{\phi}\dot{a}}{\phi a}  +  \frac{\lambda}{6\phi^{2}\sqrt{1 -
    \dot{\phi}^{2}}} \label{freq1}\\
\nonumber \\
\left(\frac{\ddot{a}}{a}\right) & =& -\frac{4\pi\phi^{2}}{3}(\rho + 3p) +
\frac{\dot{\phi}\dot{a}}{\phi a} + \frac{\ddot{\phi}}{\phi} -
3\frac{\dot{\phi}^{2}}{\phi^{2}}\nonumber\\
& & + \frac{\lambda}{4\phi^{2}}\sqrt{1 -
      \dot{\phi}^{2}} - \frac{\lambda}{12\phi^{2}}\frac{1}{\sqrt{1 -
      \dot{\phi}^{2}}} \label{freq2}
\end{eqnarray}
With $\phi = \phi(t)$ the equation for $\phi$ (Eq.(\ref{tsi})) takes the form:

\begin{eqnarray}
8\pi\phi^{2}(\rho - 3p) &=& -6\frac{\ddot{\phi}}{\phi}
-18\frac{\dot{\phi}\dot{a}}{\phi a} + 18\frac{\dot{\phi}^{2}}{\phi^{2}} +
\frac{3\lambda\dot{\phi}^{2}}{2\phi^{2}\sqrt{1 - \dot{\phi}^{2}}}\nonumber \\
& & - \frac{\lambda\ddot{\phi}}{2\phi\sqrt{1 - \dot{\phi}^{2}}} -
\frac{3\lambda\dot{a}\dot{\phi}}{2a\phi\sqrt{1 - \dot{\phi}^{2}}} -
\frac{\lambda\dot{\phi}^{2}\ddot{\phi}}{2\phi(1 -
  \dot{\phi}^{2})^{3/2}}\label{freq3}
\end{eqnarray}
Here $\rho = T_{0}^{0}$ and $p =  -T_{1}^{1} = - T_{1}^{1} = - T_{1}^{1} $
which is the energy density and the pressure, respectively.

In our model we assume the universe to consist of the pressureless matter in
addition to the field $\phi$.
The contribution from radiation at late time (as is the case for the present era)
can be neglected. Hence, we drop the contribution of radiation in the source term
for Einstein's equation. 
In addition to equations (\ref{freq1}) and (\ref{freq2}) we also have the continuity equation for matter.
\begin{equation}
\dot{\rho} = -3H(\rho + p) \label{energyC}
\end{equation}
For pressure less matter this gives:
\begin{equation}
\rho_{m}(a) = \rho_{m0}\frac{a_{0}^{3}}{a^{3}}
\end{equation}
where $ \rho_{m0}$ is the matter density at the present epoch and $ a_{0}$ is
the scale factor at the present epoch.

In order to solve these equations (Eqs.(\ref{freq1}) and (\ref{freq2}))
we need the values of $\rho_{m0}$ and $\lambda$ as well as  the values of
$a$ , $\phi$ and $\dot{\phi}$ at some instant of time.
We now fix these values from the observations at the present epoch.
First of all we normalize the scale factor to unity at the present
epoch. At the present epoch $\dot{a}(t_{0}) = H_{0}$ (as $a(t_0)=1$) which is
the Hubble constant. The value of $\phi$ at the present epoch
will be fixed as $ \phi(t_{0}) = \sqrt{G_{0}}$ where $G_{0}$ is the Newtonian
gravitational constant at the present epoch. There is strong evidence from both
experiments as well as cosmological observation that there is no significant
variation of the gravitational constant at the present epoch. These considerations
will be consistent if we consider the simplest case when $\dot{\phi}(t_{0}) = 0$
which ensures that $\dot{G}(t_{0}) = 0$. So we consider the following conditions
at the present epoch, $ t = t_{0}$ :
\begin{eqnarray}
a(t_0) &=& 1\label{at0}\\
\dot{a}(t_{0}) &=& H_{0}\\
\phi(t_{0}) &=& \sqrt{G_{0}}\\
\dot{\phi}(t_{0}) &=& 0\label{sidt0}
\end{eqnarray}

\begin{figure}
\centering{
\scalebox{2}{\rotatebox{270}{
\includegraphics[width = 1in]{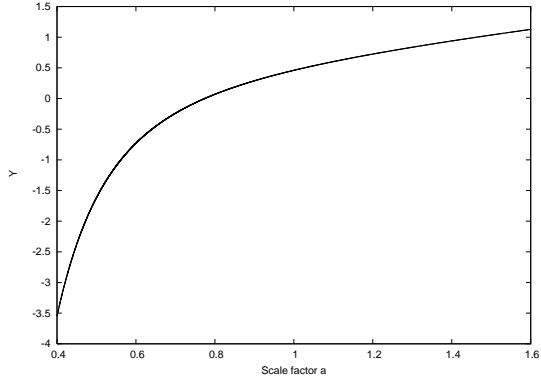}
}}}
\caption{This plot shows how $\ddot{a}$ evolves with scale factor.
Here $ Y = \ddot{a}/H_{0}^{2} $}
\label{adda}
\end{figure}

With these conditions we can determine the value of $\lambda$ from Eq.
(\ref{freq1}). On substituting Eqs. (\ref{at0}) to (\ref{sidt0}) in the
Friedmann equation (\ref{freq1}) we get:
\begin{equation}
\lambda = 6G_{0}H_{0}^{2}(1 - \Omega_{m})\label{lamda},
\end{equation}
where $\Omega_{m}$ is the density parameter of (pressureless) matter today
and is defined as
\begin{equation}
 \Omega_{m} = \frac{8\pi G_{0}\rho_{m0}}{3H_{0}^{2}}
\end{equation}
Observations  indicate that $\Omega_{m} = 0.27$ and $H_{0} =
70km/s/Mpc$ (as indicated by the observations by WMAP \cite{bennett}). In the units of $\hbar = c = 1$ this gives the value $H_{0} =
6.1207 \times 10^{-61}M_{p}$  and with the value of Newtonian gravitational constant
$G_{0} = (8\pi)^{-1}M_{p}^{-2}$ , we obtain the value of $\lambda$ as :
\begin{equation}
\lambda = 6.5281 \times 10^{-122}
\end{equation}
Eq.(\ref{freq3}) together with conditions Eqs.(\ref{at0}) to
(\ref{sidt0}) gives,
\begin{equation}
\ddot{\phi}(t_{0}) = -\frac{6H_{0}^{2}\Omega_{m}\sqrt{G_{0}}}{(12 +
  \lambda)}\label{sidd}
\end{equation}
If the time variation of the field $\phi$ is exactly zero in the
present epoch it would imply that $\ddot{G}(t_{0})$ is negative which
in turn implies that
$G(t)$ has a maxima at the present epoch. Observationally we know
that there is no significant variation  of the "Gravitational coupling"
with time. Hence, with in the framework of this scenario, the present
value of the Gravitational coupling parameter should be close to its
maximum value.
Eq.(\ref{freq2}) together with conditions Eqs.(\ref{at0}) to
(\ref{sidt0}) gives,
\begin{equation}
\frac{\ddot{a}(t_{0})}{H_{0}^{2}} = -\frac{\Omega_{m}}{2} +
\frac{\ddot{\phi}(t_{0})}{H_{0}^{2}\sqrt{G_{0}}} +
\frac{\lambda}{6H_{0}^{2}G_{0}}. \label{add}
\end{equation}
From Eqs. (\ref{lamda}) , (\ref{sidd}) and (\ref{add}) we get:
\begin{equation}
\frac{\ddot{a}(t_{0})}{H_{0}^{2}} = 1 - 2\Omega_{m} = 0.46
\label{addt0}
\end{equation}
\begin{figure}
\centering{
\scalebox{2}{\rotatebox{270}{
\includegraphics[width = 1in]{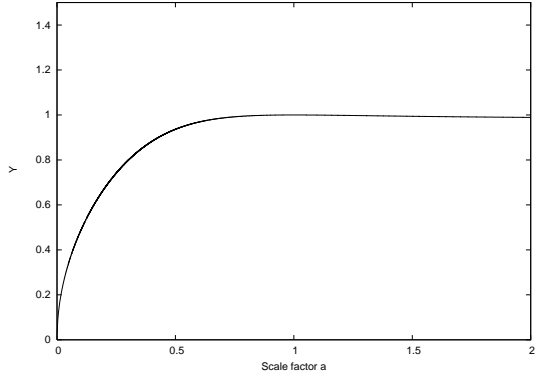}
}}}
\caption{This plot shows how field $\phi$ evolves with scale factor. Here
$Y = \phi/\sqrt{G_{0}}$}
\label{Ga}
\end{figure}
Hence the conditions Eqs.(\ref{at0}) to (\ref{sidt0}) naturally leads to an
accelerated expansion of the universe at the present epoch.
Within the frame work of Einstein GR, if we consider dark energy to be a
perfect fluid with equation of state $ p_{\phi} = w_{\phi}\rho_{\phi}$, then
such a value of $ \ddot{a}(t_{0})$ (Eq.(\ref{addt0})) would correspond to
$ w_{\phi}(t_{0}) = -0.88$. The figure (\ref{adda}) shows the plot of
$\ddot{a}/H_{0}^{2}$ verses the scale factor $a$. We see from the figure that
$\ddot{a}$ is less than zero for small values of the scale factor and becomes
positive for the large values. The transition occurs for $a=.77$
(corresponding to a redshift of $z = 0.3$). Thus in this model the transition
from a decelerating phase to an accelerating phase should have taken place
at $z=0.3$. In figure (\ref{Ga}) we plot the ratio $\phi/\sqrt{G_{0}}$ verses
the scale factor, where $G_{0}$ is the Newtonian gravitational constant.
We can see from the figure (Fig.(\ref{Ga})) that the field $\phi$
asymptotically attains a constant value of $ 0.98 \times \sqrt{G_{0}}$.
A constant value of the field $\phi$ would imply that the field equation
(Eq.(\ref{feqgab})) would then become usual Einstein equation with a cosmological
constant. Hence this model asymptotically becomes De Sitter universe.

\section{Conclusion and Discussion}    \label{conc}
We have considered a model in which the gravitational coupling evolves
as  $G \sim \phi^2$ where the Lagrangian for the scalar field, $\phi$,
is given by the Lagrangian of the form
$\mathcal{L}_{\phi} = -V(\phi)\sqrt{1 -
\partial_{\mu}\phi\partial^{\mu}\phi}$. We have chosen a potential of the
form $V(\phi) = \lambda \phi^{-4}$, where $\lambda$ is a dimensionless
parameter. On considering the cosmological consequence of this theory we have
shown that imposing condition that $\dot{\phi}(t_{0})= 0$ gives a value of
$ \lambda = 6.5281 \times 10^{-122}$ and also gives a positive value of
acceleration ( at the present epoch)  with $ \ddot{a}(t_{0}) = 0.46H_{0}^{2}$.
Such a value of $\ddot{a}$ would correspond to an effective equation of state
$w_{\phi}(t_{0}) = -0.88$. Numerical calculation shows that there was a
transition from an initial decelerating expansion to an accelerated expansion
phase at the present epoch. This transition occurred at $z = 0.3$. This result
is consistent with the observations\cite{riess,perl}. Hence this could be a
possible model for the present observed accelerated expansion of the universe. \\

\begin{flushleft}
\textbf{ Acknowledgment}\\

S.U. thanks C.S.I.R, India for a Senior Research Fellowship.
TRS thanks IUCAA for the support provided
through the Associateship Program and DST, India for the project. S.U and TRS
thanks IUCAA  for the facilities at the IUCAA Reference Centre at Delhi University.
\end{flushleft}

\begin{flushleft}
\textbf{APPENDIX}\\
\end{flushleft}
\textbf{Action for scalar tensor theory of gravitation}\\
Let us assume that the locally measured gravitational constant $G$ is
determined by the value of a scalar field $\phi$, \textit{i.e}
$ G = f(\phi)$. Action for such a theory would be :
\begin{equation}
\mathcal{A} = \frac{1}{16\pi} \int \left[-\frac{R}{f(\phi)}
+ \mathcal{L}_{\phi} +
16\pi\mathcal{L}_{m}\right] \sqrt{-g}~d^{4}x  \label{STTG}
\end{equation}
where $\mathcal{L}_{m}$ is the Lagrangian of matter and
$\mathcal{L}_{\phi}$ is the Lagrangian of the scalar field
$\phi$ (un-coupled part). Note that in Einstein-Hilbert Action the
coefficient of the Ricci scalar  $R$ is $1/16\pi G$ but here we have
$1/16\pi f(\phi)$.
In this paper we have studied the cosmological consequences of a
 scalar field with the Lagrangian of the form
$\mathcal{L}_{\phi} = -V(\phi)\sqrt{1 - \partial_{\mu}\phi\partial^{\mu}\phi}$.
We now obtain the form of $f(\phi)$ and $V(\phi)$ from the dimensional argument
but it will, however, be instructional to first consider briefly the case of a
scalar field with the Lagrangian in the standard form, namely,
$\mathcal{L}_{\phi}=\frac{1}{2}\partial_{\mu}\phi\partial^{\mu}\phi -V(\phi)$.
The dimension of $\phi$ is $M$. Since the dimensions of $G$ as well
as $f(\phi)$ is $M^{-2}$. The simplest form for the function $f(\phi)$
would be $f(\phi) = \alpha \phi^{-2}$, where $\alpha$
is a dimensionless parameter. Consider the simplest case when $V(\phi) = 0$.
With this the action takes the form :
 \begin{equation}
\mathcal{A} = \frac{1}{16\pi} \int \left[-\frac{\phi ^{2} R}{\alpha} +
\frac{1}{2}\partial_{\mu}\phi\partial^{\mu}\phi +
  16\pi\mathcal{L}_{m}\right] \sqrt{-g}~d^{4}x  \label{STTGb}
\end{equation}
Defining $\phi_{BD} = \phi^{2}/\alpha$ and $ w = \alpha / 8$ the action
(\ref{STTGb}) can be expressed in an alternate form as
\begin{equation}
\mathcal{A} = \frac{1}{16\pi} \int \left[-\phi_{BD} R +
w\frac{ \partial_{\mu}\phi_{BD}\partial^{\mu}\phi_{BD}}{\phi_{BD}} +
  16\pi\mathcal{L}_{m}\right] \sqrt{-g}~d^{4}x  \label{STTGbdt}
\end{equation}
This is the standard action for the Brans-Dicke Theory with field
$\phi_{BD}$ being the
Brans Dicke field.
\vskip 0.5 cm
\noindent
\underline{$\mathbf{\mathcal{L}_{\phi} = -V(\phi)\sqrt{1 -
\partial_{\mu}\phi\partial^{\mu}\phi}}$}  :
\vskip .25 cm
In this case the dimension of scalar field $\phi$ is $M^{-1}$.
As the dimension of $f(\phi)$ is $M^{-2}$ the simplest choice from
dimensional argument would be $f(\phi) = \phi^{2}$.
The dimension of potential $V(\phi)$ is $M^{4}$. So the simplest choice would
be to consider $V(\phi) = \lambda \phi^{-4}$, where $\lambda$ is a
dimensionless parameter. With this the action (\ref{STTG}) becomes
\begin{equation}
\mathcal{A} = \frac{1}{16\pi} \int \left[-\frac{R}{\phi^{2}} -
\frac{\lambda}{\phi^{4}}\sqrt{1 - \partial_{\mu}\phi\partial^{\mu}\phi} +
  16\pi\mathcal{L}_{m}\right] \sqrt{-g}~d^{4}x  \label{STTGtc}
\end{equation}
This is the form of action that we have considered in this paper.

[It is interesting to note that the action in Eq.(\ref{STTGtc}) is related
to Brans-Dicke action in the limit
$\partial_{\mu}\phi\partial^{\mu}\phi << 1 $. This limit is
reminiscent of the
non-relativistic limit in particle mechanics where we use  $\dot{x} \ll c$.
In this limit the action
Eq.(\ref{STTGtc}) reduces to the form.
 \begin{equation}
\mathcal{A} = \frac{1}{16\pi} \int \left[-\frac{R}{\phi^{2}}  +
  \frac{\lambda}{2\phi^{4}} \partial_{\mu}\phi\partial^{\mu}\phi -
  \frac{\lambda}{\phi^{4}} +  16\pi\mathcal{L}_{m}\right] \sqrt{-g}~d^{4}x
\end{equation}
Denoting $\phi_{BD} = \phi^{-2}$ and $w = \lambda/8$ , this action can be
expressed in an alternate form
\begin{equation}
\mathcal{A} = \frac{1}{16\pi} \int \left[-\phi_{BD} R +
w\frac{ \partial_{\mu}\phi_{BD}\partial^{\mu}\phi_{BD}}{\phi_{BD}} -
8w\phi_{BD}^{2} +  16\pi\mathcal{L}_{m}\right] \sqrt{-g}~d^{4}x
\end{equation}
This is the action for Brans-Dicke theory with a quadratic potential for the
Brans-Dicke
field $\phi_{BD}$.  Such a quadratic form of the potential in Brans-Dicke
theory has
been considered earlier in literature \cite{martins}.]

\end{document}